\documentclass[prl,showpacs,floatfix,twocolumn]{revtex4}

\usepackage{makeidx}
\usepackage{amssymb}
\usepackage{amsmath}
\usepackage{amsfonts}
\usepackage{graphicx}

\begin{document}

\title{Entangling capacities of noisy two-qubit Hamiltonians}
\author{Somshubhro Bandyopadhyay}
\affiliation{Chemical Physics Theory Group, Department of Chemistry, 80 St. George St.,
University of Toronto, Toronto, ON, M5S 3H6, Canada}
\author{Daniel A. Lidar}
\affiliation{Chemical Physics Theory Group, Department of Chemistry, 80 St. George St.,
University of Toronto, Toronto, ON, M5S 3H6, Canada}

\begin{abstract}
We show that intrinsic fluctuations in system control parameters impose
limits on the ability of two-qubit (exchange) Hamiltonians to generate
entanglement starting from mixed initial states. We find three 
classes for Gaussian and Laplacian fluctuations. For the Ising and XYZ
models there are qualitatively distinct sharp entanglement-generation
transitions, while the class of Heisenberg, XY, and XXZ Hamiltonians is
capable of generating entanglement for any finite noise level. Our findings
imply that exchange Hamiltonians are surprisingly robust in their ability to
generate entanglement in the presence of noise, thus potentially reducing
the need for quantum error correction.
\end{abstract}

\pacs{03.67.Lx, 05.30.Ch, 03.65.Fd}
\maketitle

Considerable experimental efforts have been devoted in the past few years to
the creation of entangled states, with impressive success in systems such as
trapped ions, coupled atomic gas samples, polarized photons, and most
recently, superconducting qubits \cite{Sackett:00-Julsgaard:01-Eibl:03-Pashkin:03Berkley:03}. An important
motivation comes from quantum information processing (QIP), where
entanglement is believed to play an important role in algorithmic speedup,
communication tasks, and cryptographic applications \cite{Nielsen:book}. As
the generation of entanglement often involves the manipulation of an
interaction Hamiltonian, recent theoretical work has focused on the
entanglement capabilities of such Hamiltonians. In particular, questions
concerning optimality \cite{Dur:01,Dodd:01Wocjan:01}, equivalence classes 
\cite{Zhang:03}, and entangling power/capacity \cite{Zanardi:00Childs:02Wang:02}, have been raised and answered, under the
assumption of \emph{noiseless} controls. Here we take the first step toward
addressing what happens when this assumption is relaxed. In particular, we
wish to find out the answer to the following question: \emph{What are the
limits imposed on entanglement generation via two-body Hamiltonians by
fluctuations in system-control parameters?} \cite{comment3-NH}. We note
that, as is well known, quantum error correction \cite{Nielsen:book} offers
a solution to both decoherence and the type of control errors we consider
here; however, this solution involves a high cost in extra qubits and logic
gates. In view of the central importance of entanglement in QIP, it is of
significant interest to find out the limits imposed on entanglement
generation via interaction Hamiltonians and state preparation, without any
error correction.

\textit{The Model}.--- Almost all quantum computing proposals are governed
by interaction Hamiltonians that are used to enact two qubit operations. The
most general two-qubit (\textquotedblleft exchange\textquotedblright )
Hamiltonian has the form 
\begin{equation}
H=\sum_{i<j}\sum_{\alpha ,\beta =x,y,z}J_{\alpha \beta }^{ij}S_{\alpha}^{i}S_{\beta }^{j}
\end{equation}
where $S_{\alpha }\equiv \frac{1}{2}\sigma _{\alpha }$ are the angular
momentum operators ($\sigma _{\alpha }$ are the Pauli matrices), and $i,j$
are qubit indices. By applying local unitary operations it is always
possible to transform $H$ into a canonical, diagonal form \cite{Dur:01}.
Hence we will restrict our attention to the case $J_{\alpha \beta}^{ij}=J_{\alpha }^{ij}\delta _{\alpha \beta }$ from now on. The various
models are then classified as follows: XYZ: $J_{x}\neq J_{y}\neq J_{z}$,
XXZ: $J_{x}=J_{y}\neq J_{z}$, XY: $J_{x}=J_{y},J_{z}=0$, Heisenberg: $
J_{x}=J_{y}=J_{z}$, Ising: $J_{x}=J_{y}=0$.

Two qubits can be entangled by first preparing a product state and then
running the interaction for a desired amount of time, to generate, e.g., a
C-NOT or C-PHASE gate \cite{Nielsen:book}. Tunability of the coupling
constants in $H$ need not always be possible, even though it is a common
assumption of QIP proposals. This leads to two qualitatively distinct
scenarios we consider in this work: (a) Tunable interactions -- where the
interaction can be switched on and off (e.g., an exchange interaction
mediated by a tunable tunneling barrier \cite{Burkard:99}); (b) Non-tunable
interactions -- where the interaction is always on (e.g., a Coulomb
interaction \cite{Dykman:00}), thus requiring, e.g., external single qubit
operations to refocus the interactions and enable controlled entanglement
generation. Recent work has addressed the problem of universal quantum
computation with non-tunable couplings \cite{Zhou:02}, and even unknown
parameters \cite{Garcia-Ripoll:02}. In the laboratory, however, the
execution of every single and two qubit operation will generally be noisy
due to system and experimental imperfections, over which we have
limited control. We consider phenomenological noise models to
describe noisy single qubit and two qubit operations, wherein certain
control parameters vary stochastically. Specifically, we have considered two
models: (a) Gaussian and (b) Laplacian parameter fluctuations. The Gaussian
model has universal applicability in the case of noise due to many weakly
coupled random sources (by the central limit theorem). It has been
extensively used and discussed in stochastic quantum mechanics (e.g., \cite
{Cheng:03}). We consider the Laplacian model mainly to test the robustness
of our results. Another important model is $1/f$ noise due to bistable
random fluctuators, which will be considered in a future publication.

\textit{Noise model}.--- Given a Hamiltonian $K(t,J)$ (where $J$ is a
parameter or set of parameters), a unitary transformation $U$ is generated
by evolving under $K$ for some time $\tau $: $U(\phi )=\mathcal{T}\exp
(-i\int_{0}^{\tau }K(t,J)dt)$, where $\mathcal{T}$ denotes time ordering. In
our analysis below we only deal with piece-wise constant Hamiltonians; then
the angle(s) $\phi =\tau J$. An initial state $\rho $ transforms as $\rho
\rightarrow \rho (\phi )=U(\phi )\rho U(\phi )^{\dag }$. We now assume that $
\phi $ is Gaussian distributed with mean $\bar{\phi}$ (the desired angle)
and standard deviation (s.d.) $\lambda $: $\phi \sim N(\bar{\phi},\lambda )$. This may be the outcome of Gaussian noise in $J$, $\tau $, or both. Thus,
under a noisy control the actual transformation is 
\begin{equation}
\rho \rightarrow \rho _{\mathrm{noisy}}\left( \lambda ,\bar{\phi}\right) =
\frac{1}{\sqrt{2\pi }\lambda }\int_{-\infty }^{\infty }e^{-\frac{(\phi -\bar{
\phi})^{2}}{2\lambda ^{2}}}\rho \left( \phi \right) d\phi ,  \label{eq:noisy}
\end{equation}
Below we take $K$ to be either an exchange Hamiltonian with noisy coupling
constants $J_{\alpha }^{ij}$ (in which case we assume for simplicity an
equal s.d., denoted $\Omega $), or a noisy single-qubit Hamiltonian needed
to refocus an always-on exchange Hamiltonian (and denote the s.d. by $\Lambda $).

\textit{Case (i): Tunable Ising interaction}.--- In the Ising model, the
exchange Hamiltonian takes the form $H_{ZZ}=\frac{1}{4}J\sigma
_{z}^{1}\sigma _{z}^{2}=JS_{z}^{1}S_{z}^{2}$. Consider the preparation of a
maximally entangled state, starting from the initial state $\rho
_{i}=\left\vert 00\right\rangle \left\langle 00\right\vert $, in the
presence of noisy interactions \cite{comment1-NH}. Without noise,
application of the Hadamard transform $U_{H}(\pi )=\exp (-i\pi S_{y})$ on
both qubits, followed by $U_{ZZ}(\pi )=\exp (-i\int_{0}^{\tau }\frac{1}{4}
J\sigma _{z}^{1}\sigma _{z}^{2}dt)=e^{-i\frac{\pi }{4}\sigma _{z}^{1}\sigma
_{z}^{2}}$, prepares the maximally entangled pure state $\left\vert \xi
\right\rangle =\frac{1}{2}\left( \left\vert 00\right\rangle +i\left\vert
01\right\rangle +i\left\vert 10\right\rangle +\left\vert 11\right\rangle
\right) $. In the presence of noise, the action of both $U_{H}$ and $U_{ZZ}$
must be averaged over a distribution of angles, as in Eq.~(\ref{eq:noisy}).
The noisy Hadamard transform is a rotation about the $y$-axis with average
angle $\pi $ and s.d. $\lambda $, resulting in the mixed state $\rho
_{0}\left( \lambda \right) =\frac{1}{\sqrt{2\pi }\lambda }\int e^{-\frac{
(\theta -\pi )^{2}}{2\lambda ^{2}}}\left( U_{H}(\theta )\rho
_{i}U_{H}(\theta ){}^{\dagger }\right) d\theta $, which can be easily
evaluated. The mixedness of $\rho _{0}$ is measured by its von Neumann
entropy: $M\left( \rho \right) =-\mathrm{Tr}[\rho \log _{2}\rho ]$, as a
function of the noise parameter $\lambda $. $M\left( \rho \right) =0$ for a
pure state; $M\left( \rho \right) =2$ for a maximally mixed state. In the
present case the entropy rises rapidly from zero (at $\lambda =0$) and
reaches its maximum of $2$ for $\lambda \approx 2$. Next we apply the noisy
version of the $U_{ZZ}$ gate, with angle $J\tau \sim N(\pi ,\Omega )$. The
resulting density matrix $\rho \left( \lambda ,\Omega \right) $ is easily
computed but is not particularly illuminating; instead we present the result
of using the partial transposition test for entanglement \cite{Peres:96Horodecki:96}: a $2\otimes 2$ state is entangled iff it has
negative partial transpose (NPT). One then arrives at the following
condition for inseparability for the density matrix $\rho \left( \lambda,\Omega \right) $ 
\begin{equation}
e^{-\lambda ^{2}}+2e^{-\frac{1}{2}(\lambda ^{2}+\frac{\Omega ^{2}}{4})}>1
\label{eq:ZZinsep}
\end{equation}

\begin{figure}[tbp]
\includegraphics[height=5cm,angle=0]{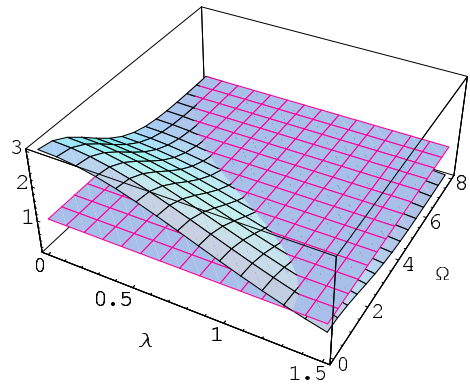}
\caption{Condition for entanglement as a function of interaction error $
\Omega $ and preparation error $\protect\lambda $ in the Ising model.
Plotted is the PPT criterion~(\protect\ref{eq:ZZinsep}). The transition line
between entanglement and separability is clearly visible, with the region
above the horizontal plane correponding to entanglement.}
\label{fig2}
\end{figure}
\noindent Figure~\ref{fig2} illustrates this condition. Observe that there
is a significant region of entanglement in parameter space, with a trade-off
between the tolerated level of noise in state preparation and interaction.
Interestingly, except along the cut $\lambda =0$, \emph{the transition from
entangled to separable is sharp}. Solving the inequality~(\ref{eq:ZZinsep}),
we find that the condition for entanglement is 
\begin{equation}
\lambda \leq \{-2\log [(e^{-\frac{\Omega ^{2}}{4}}+1)^{1/2}-e^{-\frac{\Omega
^{2}}{8}}]\}^{1/2}.  \label{eq:ZZcond}
\end{equation}
The finite range of the state preparation parameter $\lambda $ indicates
that \emph{the purity of the initial state is crucial}. However, with a high
quality interaction a significantly mixed initial state can be tolerated:
From the above follows that if the interaction is perfect ($\Omega =0$) then 
$\lambda _{\mathrm{max}}\allowbreak =\allowbreak (-2\ln \left( \sqrt{2}
-1\right) )^{1/2}=\allowbreak 1.\,\allowbreak 327$, meaning that even an
initial mixed state with entropy $96\%$ as high as the maximally mixed state
would still enable entanglement generation. Conversely, if the initial state
preparation does not involve any noise ($\lambda =0$), then the interaction,
no matter how noisy will be able to produce some entanglement. Finally, note
that for $2\otimes 2$ systems of the type we are considering here, if a
state is entangled then it is distillable as well, i.e., one can extract
pure states from such noisy states using local operations and classical
communication \cite{Horodecki:97}. \emph{In particular}, Eq.~(\ref{eq:ZZcond}) \emph{therefore guarantees that a state is useful for} teleportation \emph{
and all other QIP primitives}.

\textit{Case (ii): Untunable Ising interaction}.--- Now we do not assume the
ability to switch the interaction off (as, e.g., in NMR). It is therefore
necessary to refocus the interaction using single qubit operations \cite{Jones:99}. This is done by pulsing an external magnetic field along the $x$
-axis (we choose the first qubit for this operation).{} We assume that such
pulses can be made very fast and strong compared to the interaction \cite{Jones:99}. Formally, let $X,Y,Z$ be operators satisfying $su(2)$
commutation relations: $[X,Y]=iZ$ and cyclic permutations (e.g., the
angular momentum operators $S_{\alpha }$). Then it follows from the
Baker-Campbell-Hausdorff formula that upon
\textquotedblleft conjugation by $\varphi $\textquotedblright : $
C_{Z}^{\varphi }\circ X\equiv \exp (-i\varphi Z)X\exp (i\varphi Z)=X\cos
\varphi +Y\sin \varphi .$ Note further that $Ue^{A}U^{\dagger
}=e^{UAU^{\dagger }}$ for unitary $U$ and arbitrary $A$. Thus: $
C_{Z}^{\varphi }\circ \exp (i\theta X)=\exp (-i\varphi Z)\exp (i\theta
X)\exp (i\varphi Z)=\exp (i\theta (X\cos \varphi +Y\sin \varphi ))$. One
application is \textquotedblleft time-reversal\textquotedblright , which
results from conjugation by $\pi $:

\begin{equation}
C_{Z}^{\pi }\circ e^{i\theta X}=e^{-iZ\pi }e^{i\theta X}e^{iZ\pi
}=e^{-i\theta X}.
\end{equation}
We then have $\rho (\tau _{2})=U_{R}(\pi )\rho _{0}U_{R}(\pi )^{\dagger }$ ,
where $U_{R}(\pi )=[C_{S_{x}^{1}}^{\pi }\circ \exp (-i\int_{\tau _{1}}^{\tau
_{2}}JS_{z}^{1}S_{z}^{2}dt)]\exp (-i\int_{0}^{\tau
_{1}}JS_{z}^{1}S_{z}^{2}dt)=e^{-i\frac{\pi }{4}\sigma _{z}^{1}\sigma
_{z}^{2}}$, which in conjunction with our Hadamard state preparation yields
the desired C-PHASE gate. The parameters must satisfy the condition $J(2\tau
_{1}-\tau _{2})=\pi $. In the noisy scenario the recoupling step is
implemented with a rotation around the $x$ axis by an angle $\theta \sim
N(\pi ,\Lambda )$ (where $\Lambda^2=\Lambda_1^2+\Lambda_2^2$, with
$\Lambda_i$ the s.d.'s of the independent Gaussian random variables
$J\tau_i$, $i=1,2$, $J$ fixed); the only change is then $U_{R}(\pi )\mapsto U_{R}(\theta )
$, and the final state is given by: $\rho (\tau _{2},\lambda ,\Lambda )=
\frac{1}{\sqrt{2\pi }\Lambda }\int e^{-\frac{(\theta -\pi )^{2}}{2\Lambda
^{2}}}\left( U_{R}(\theta )\rho _{0}U_{R}(\theta )^{\dagger }\right) d\theta $, which can be analytically evaluated. The inseparability condition,
obtained using the partial transposition criterion, yields the same form as
the tunable case [Eq.~(\ref{eq:ZZinsep})], provided we replace $\Omega /2$
by $\Lambda $. Note that this is by no means an \textit{a priori} obvious
substitution: $\Omega $ is the two-body interaction strength error, whereas $
\Lambda $ is the error in the single-body coupling parameter.

\begin{figure}[tbp]
\includegraphics[height=7cm,angle=0]{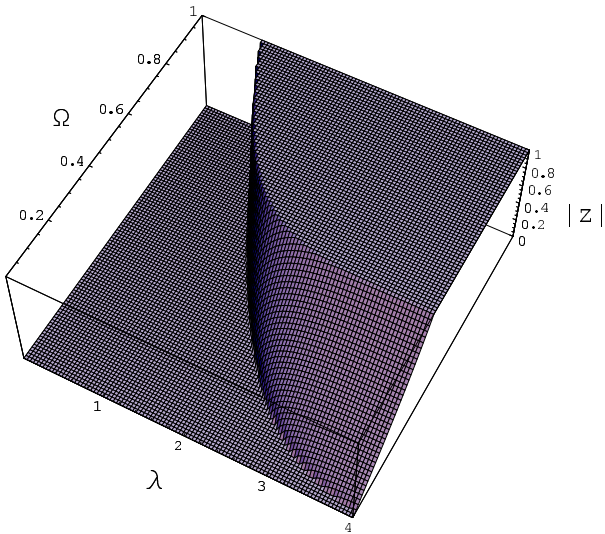}
\caption{Plot of the inseparability condition~(\protect\ref{eq:insep-cond}).
Entanglement in the noisy tunable XYZ model is the region where $|z|$ is
above the surface shown and $<1$.}
\label{fig:XYZ}
\end{figure}

\textit{Case (iii): Tunable XYZ Hamiltonians}.--- We now consider the XYZ
model $H_{\mathrm{XYZ}}=\sum_{\alpha =x,y,z}J_{\alpha }S_{\alpha
}^{1}S_{\alpha }^{2}$, of which, \emph{in the noiseless case}, the XXZ, XY
and Heisenberg models are special cases. As in the Ising case we assume the
initial state is $\rho _{i}=\left\vert 00\right\rangle \left\langle
00\right\vert $ \cite{comment2-NH}. In the noiseless scenario, we first
apply $U_{X}(\pi )=\exp (-i\pi S_{x})$ on the second qubit, yielding $\rho
_{0}=\left\vert 01\right\rangle \left\langle 01\right\vert $. This is
followed by $U_{\mathrm{XYZ}}(\theta _{x},\theta _{y},\phi )=\exp
(-i\int_{0}^{\tau }H_{\mathrm{XYZ}}dt)$, where $\theta _{x,y}=J_{x,y}\tau $, 
$\phi =J_{z}\tau $. Letting $\theta _{x}+\theta _{y}=\frac{\pi }{4}$ \emph{
and leaving $\phi $ arbitrary}, this prepares the maximally entangled pure
state $\left\vert \xi \right\rangle =\frac{1}{\sqrt{2}}\left( i\left\vert
01\right\rangle +\left\vert 10\right\rangle \right) $. Thus in the noiseless
case, and for all the exchange models considered here, there is no
dependence on $J_{z}$. In the noisy scenario, we first apply $U_{X}(\omega)=\exp (-i\omega S_{x})$ where $\omega \sim N(\pi ,\lambda )$. Then $\rho
_{0}\left( \lambda \right) $ is a mixture of the states $\left\vert
00\right\rangle $,$\left\vert 01\right\rangle $ with respective weights $
[1\mp \exp (-\frac{\lambda ^{2}}{2})]/2$. Note that now, in the worst case
scenario ($\lambda \rightarrow \infty $) $\rho _{0}$ can only be $50\%$ as
mixed as the totally mixed state, whereas in the Ising model above $\rho _{0}$ could be fully mixed (or $96\%$ mixed if we insist on entanglement
generation).

The unitarily transformed density matrix is given by $\rho _{0}\left(
\lambda \right) \mapsto U_{\mathrm{XYZ}}\left( \theta _{x},\theta _{y},\phi
\right) \rho _{0}\left( \lambda \right) U_{\mathrm{XYZ}}^{\dagger }\left(
\theta _{x},\theta _{y},\phi \right) =\rho (\lambda ,\theta ^{+},\theta
^{-}) $, where $\theta ^{\pm }=\theta _{x}\pm \theta _{y}$, whose explicit
form can be found without much difficulty, and again is independent of $\phi 
$ (i.e., $J_{z}$), now even when the initial state is noisy. We next
integrate over $\theta ^{+}\sim N(\pi /4,\Omega )$ and $\theta ^{-}\sim N( 
\overline{\theta ^{-}},\Omega )$, where $\Omega = (\Omega_{x}^{2}+\Omega
_{y}^{2})^{1/2} $, and where $\Omega _{i}$ are the s.d.'s of $\theta _{i}$.
Upon applying the partial transposition criterion we find the condition for
inseparability

\begin{equation}
|z|>[1/b^{2}-(1+a)^{2}/(1-a)^{2}]^{1/2},  \label{eq:insep-cond}
\end{equation}
where $a=e^{-\frac{1}{2}\lambda ^{2}}$, $b=e^{-2\Omega ^{2}}$, $z=\cos (2
\overline{\theta ^{-}})$. This condition, plotted in Fig.~\ref{fig:XYZ},
depends on the noise parameters $\lambda ,\Omega $ (as in the Ising case),
but also periodically on the (mean) coupling constants through the distance
between exchange models $\overline{\theta ^{-}}=(\overline{J_{x}-J_{y}})\tau $. From $0\leq |z|,a,b\leq 1$, we have the following sufficient conditions
for inseparability [RHS of Eq.~(\ref{eq:insep-cond}) $<0$]: $b>(1-a)/(1+a)$
(curve at $|z|=0$ in Fig.~\ref{fig:XYZ}), and separability [RHS of Eq. (\ref
{eq:insep-cond}) $>1$]: $b<(1-a)/\sqrt{2\left( 1+a^{2}\right) }$ (curve at $
|z|=1$ in Fig.~\ref{fig:XYZ}). Thus we find that that \emph{the XYZ model
with} $J_{x}\neq J_{y}$, similarly to the Ising model, \emph{exhibits a
sharp entanglement/separability transition as a function of preparation and
interaction noise}. However, in contrast to the Ising model, in general
there is also a dependence on a third parameter, the (noisy) interaction
distance $J_{x}-J_{y}$, so the XYZ model belongs to a distinct 
class.

\textit{Case (iv): Tunable XY, XXZ and Heisenberg Hamiltonians}.--- Note
that because of the integration over the distribution of $J_{x}-J_{y}$ above
we can\emph{not} specialize to the XY, XXZ, Heisenberg models ($J_{x}\equiv
J_{y}=:J$ case). Repeating the calculations above now with $J\sim N(\frac{
\pi }{4},\Omega )$ we find that the final state $\rho (\lambda ,\Omega )$ is
a mixture of the states $|00\rangle $ and $\left\vert \Psi ^{\pm
}\right\rangle =\frac{1}{\sqrt{2}}\left( \left\vert 01\right\rangle \pm
\left\vert 10\right\rangle \right) $, with respective weights $\frac{1}{2 }(
1-e^{-\lambda ^{2}/2}) $ and $\frac{1}{4}( 1+e^{-\frac{1}{ 2}\lambda ^{2}})
( 1\mp e^{-\Omega ^{2}/2}) $. Such states are always entangled as long as
the proportions of the states $\left\vert \Psi ^{\pm }\right\rangle $ are
different, which clearly holds in our case for $\lambda,\Omega <\infty $. 
\emph{Hence the final state in the tunable XY, XXZ, and Heisenberg models is
always entangled for all practical purposes}. This then is a third class.

\textit{Case (v): Untunable XYZ, XY, XXZ and Heisenberg Hamiltonians}.---
Since the interaction is always on, we need to apply external single qubit
refocusing operations. This is done by pulsing an external magnetic field
along the $z$-axis (we choose the first qubit for this operation). After a
lengthy calculation we find from the partial transposition criterion, for
the XYZ model, the inseparability condition 
\begin{equation}
\left( A+B\mu \right) ^{2}\left( 1-\eta \right) ^{2}\left( C\mu -D\right)
^{2}-\frac{1}{4}\mu ^{4}\left( 1+\eta \right) ^{2}<0
\label{eq:XYZ-untunable}
\end{equation}
where $\mu =e^{-\Lambda ^{2}/2},\eta =e^{-\frac{1}{2}\lambda ^{2}}$, $A=\cos
\beta \cos \delta ,B=\sin \beta \sin \delta ,C=\sin \beta \cos \delta,D=\cos \beta \sin \delta $, $\beta =\Delta \pi /2$,$\delta =\Delta \pi /4$,
and $\Delta =\frac{J_{x}-J_{y}}{J_{x}+J_{y}}$. The XY, XXZ and Heisenberg
models ($J_{x}=J_{y}$) corresponds to $B=C=D=0$. Interestingly not only this
is achieved when $J_{x}=J_{y}$ but XYZ mimicks XY, XXZ and Heisenberg also
when $\Delta =4n$, $n$ an integer. It can be shown that the inequality~(\ref{eq:XYZ-untunable}) is satisfied as long as $\Lambda ,\lambda <\infty $. 
\emph{Hence in all these exchange models the final state is always entangled
for all practical purposes}. In contrast to the tunable case, the untunable
XYZ, XY, XXZ and Heisenberg Hamiltonians all lie in the same class.

\textit{Laplacian fluctuations}.--- As a test of the robustness of our
Gaussian-model based conclusions, we have repeated the above analysis when
the flucutuations in control parameters obey a Laplace distribution: under a noisy control the actual state transformation is now
\begin{equation}
\rho \rightarrow \rho _{\mathrm{noisy}}\left( \omega ,\bar{\phi}\right) =
\frac{1}{4\omega }\int_{-\infty }^{\infty }e^{-\frac{|\phi -\bar{\phi}|}{
2\omega }}\rho \left( \phi \right) d\phi .
\end{equation}
While we find quantitative differences compared to the Gaussian model, the
qualitative behavior is identical, thus bolstering the universality of the
classes obtained. To illustrate this fact let us reconsider the
Ising case. Let $\omega =\lambda $ ($\Omega $) be the s.d. of
the noisy parameter controlling state preparation (interaction) in both the
tunable non-tunable case. We find, repeating the procedure above, that the
condition of inseparability can be given for \emph{both} the tunable and
non-tunable cases as

\begin{equation}
4\lambda ^{2}(\Omega ^{2}+2\lambda ^{2}+2\lambda ^{2}\Omega ^{2})<1.
\label{eq:L-ineq}
\end{equation}
Recall that, similarly, in the Gaussian case the inseparability condition of
the tunable case can be obtained by appropriate rescaling of the s.d.
Further, note that as in the Gaussian model, when the initial state is
perfect ($\lambda =0$) the state remains inseparable for all practical
purposes. By solving inequality (\ref{eq:L-ineq}) the inseparability
condition can be written as $\lambda <\frac{1}{2}[((1+\left( 1+\Omega
^{2}\right) ^{2})^{1/2}-\Omega ^{2})/\left( 1+\Omega ^{2}\right) ]^{\frac{1}{
2}}$, which shows that, again as noticed in the Gaussian case, the purity of
the intial state is crucial and actually even more so in the Laplace case.
Indeed, if the interaction is perfect , we obtain the inseparability
condition $\lambda <\sqrt[4]{1/8}=0.5946$, in contrast to the Gaussian
case where the the threshold value of $\lambda $ in the no-noise interaction
scenario is as high as $1.327$.

\textit{Conclusions}.--- 
The sharp transition
found in the Ising and XYZ models is reminiscent of the thermal entanglement
transition \cite{Arnesen:01}, and suggests an interesting avenue for further
research. The surprising robustness of entanglement to noise bodes well for
quantum information processing with reduced demands on error correction.
Another interesting implication of our work concerns \emph{entanglement
verification}: Knowing the underlying two-body interaction Hamiltonian and
corresponding level of control, an experimentalist can confidently
characterize the degree of entanglement his/her system can generate, without
needing to perform a direct, and often difficult, measurement of
entanglement \cite{Sackett:00-Julsgaard:01-Eibl:03-Pashkin:03Berkley:03}.

\textit{Acknowledgements}.--- D.A.L. gratefully acknowledges the Alfred P.
Sloan Foundation for a Research Fellowship and the DARPA-QuIST program (managed by AFOSR under agreement No.
F49620-01-1-0468) for financial support.


\begin{thebibliography}{99}
\bibitem{comment1-NH} One may consider starting instead from an arbitrary,
but known pure product state. Then the Hadamard transform on each qubit will
be replaced by a general separable two-qubit unitary, that will depend on
six Euler angles. As long as all these angles are Gaussian distributed with
the same s.d., our analysis carries through unchanged.
\bibitem{comment2-NH} Comment \cite{comment1-NH} still holds, with obvious
modifications. Note, however, that in general, if the Euler angles have
different s.d.'s, the classes will split.
\end{thebibliography}
\end{document}